\documentclass[12pt,a4paper]{article}
\usepackage{amsmath}
\usepackage{amssymb} 
\usepackage{graphicx} 
\usepackage{float}
\restylefloat{table}
\begin{document}
\begin{titlepage} 
\begin{flushright} IFUP--TH/2016\\ 
\end{flushright} 
\vskip .8truecm 
\begin{center} 
\Large\bf On the solution of Liouville equation
\end{center}
\vskip 1.2truecm 
\begin{center}
{Pietro Menotti} \\ 
{\small\it Dipartimento di Fisica, Universit{\`a} di Pisa}\\ 
{\small\it 
Largo B. Pontecorvo 3, I-56127, Pisa, Italy}\\
{\small\it e-mail: pietro.menotti@unipi.it}\\ 
\end{center} 
\vskip 0.8truecm
\centerline{ October 2016}
                
\vskip 1.2truecm
                                                              
\begin{abstract}
We give a short and rigorous proof of the existence and
uniqueness of the solution of Liouville equation with sources, both
elliptic and parabolic, 
on the sphere and on all higher genus compact Riemann surfaces.
\end{abstract}

\end{titlepage}
 
\eject

\section{Introduction}

Liouville theory plays a key role in several chapters of conformal field
theory. The first studies on Liouville theory go back in addition
to Liouville himself \cite{liouville}, to Picard
\cite{picard1,picard2} and Poincar\'e \cite {poincare}.

When only elliptic singularities are present 
the existence of the solution of the equation 
\begin{equation}\label{simpleliouville}  
\Delta\phi=e^\phi
\end{equation}  
was first given by Picard in \cite{picard1,picard2} where also the
uniqueness of the solution was proven. Picard's method does not apply
in presence of parabolic singularities.

The existence and uniqueness proof was extended to the presence of
parabolic singularities by Poincar\'e \cite{poincare}. Lichtenstein
\cite{lichtenstein} reformulated the problem of finding the solution
of the equation as a variational problem proving again the
existence and uniqueness of the solution in presence of both elliptic
and parabolic singularities.
For the case of elliptic singularities 
McOwen \cite{mcowen} and Troyanov \cite{troyanov}, also employing a
variational procedure,
proved more general results using Sobolev spaces techniques.

The proofs given in the quoted papers are lengthy due in part
to the greater generality of the problem addressed by mathematicians
(formulation on metric surfaces, achievement of some prescribed
curvature function, etc.).

The aim of this paper is to give a short and at the same time
rigorous proof of the
existence and uniqueness of the solution of Liouville equation 
with sources both elliptic and parabolic, on the sphere and all higher
genus compact Riemann surfaces which is the usual setting in conformal
theory both classical and quantum.
Obviously the proofs of the existence and uniqueness of the solution
are essential in both classical
and quantum conformal theories and thus in may be useful to have at
hand a short self-contained proof in the setting of conformal
theory. This will be achieved by combining and adapting some ideas and
methods appearing in \cite{lichtenstein,mcowen,troyanov} as we shall see 
in the following.

We shall employ a useful and non trivial decomposition given in the
paper by Lichtenstein \cite{lichtenstein} of the Liouville field in a
regular and a singular part which allows to reduce the problem to a
well defined variational problem. 
Inequalities obtained by the Lichtenstein
decomposition restrict the region in which the
minimum of the action has to be found to a bounded set. Then employing
a well known compactness criterion one finds a solution as a minimum
of the action. Uniqueness is rather easily proven.

We give first the treatment on the sphere; the extension to all
higher genus compact Riemann surfaces, employing the standard polygon
representation of the compact Riemann surfaces, does not present
difficulties and is given in Section 3.
 
The sources are introduced by imposing on the field $\phi$ the boundary
conditions
\begin{equation}\label{ellipticcondition}  
\phi+2\eta_K \log|z-z_K|^2={\rm bounded},
~~~~\eta_K<\frac{1}{2}
\end{equation}
in finite disks around the elliptic singularities $z_K$ and
\begin{equation}\label{paraboliccondition}    
\phi+\log|z-z_P|^2+\log\log^2|z-z_P|^2
={\rm bounded}
\end{equation} 
in finite disks around the parabolic singularities $z_P$ and
for the sphere ($g=0$)
\begin{equation}\label{spherecond}    
\phi+2\log |z|^2={\rm bounded}
\end{equation}
outside of a circle of sufficiently large radius. 
For $g = 1$ one
imposes in addition to the conditions
(\ref{ellipticcondition},\ref{paraboliccondition}) 
periodic
boundary conditions on the boundary of the fundamental parallelogram 
and for $g\geq 2$ periodicity of $e^\phi dz\wedge d\bar z$ on the boundary of 
the fundamental $2g$-gon in the upper $z$- plane describing the
compact Riemann surface.

\section{Existence}

We shall first deal with the sphere topology.  To $z_K$ we associate
non overlapping disks of radius $r_K$ excluding the other
singularities and $R$ is chosen such that the disk of radius $R$
contains all singularities and the previously described disks.

When only elliptic singularities are present 
we construct following Lichtenstein \cite{lichtenstein} 
a smooth positive function $\beta$ such that
in the above described disks we have
\begin{equation}\label{betabound1}
0<\lambda_m<\beta|z-z_K|^{4\eta_K}<\lambda_M
\end{equation}
and for $|z|>R$
\begin{equation}\label{betabound2} 
0<\lambda_m<\beta|z|^4<\lambda_M
\end{equation}
and elsewhere 
\begin{equation}\label{betabound3}
0<\lambda_3<\beta<\lambda_4~.
\end{equation}
Note that $\int\beta d^2z <\infty$. In addition $\beta$ 
will be normalized as to have
\begin{equation}\label{choiceofalpha} 
-\sum_K 2\eta_K+\frac{1}{4\pi}\int \beta(z')d^2z' = -2  
\end{equation}  
which is possible due to the topological restriction (see section 3) 
$\sum_K 2\eta_K >2(1-g)$
where the sum extends to the sources.
  
We define
\begin{equation}\label{v}
v= \phi_1 
+\frac{1}{4\pi}\int\log|z-z'|^2 \beta(z')d^2z'
\end{equation}
with
\begin{equation}
\phi_1 =\sum_K (-2\eta_K)\log|z-z_K|^2
\end{equation}
and $U$ by
\begin{equation}
\phi = v+U~.
\end{equation}
As $\phi$ behaves $-2\log |z|^2$ at infinity, 
$U$ behaves as a constant at infinity.
The Liouville equation (\ref{simpleliouville}) becomes in 
$C\backslash\{z_K\}$
\begin{equation}\label{newliouville}
\Delta U+\beta = e^\phi= e^v e^U\equiv r\beta e^U
\end{equation}
where we defined $e^v = r \beta$ and from the above we have that
\begin{equation}\label{rbounds}
0<\lambda_1<r<\lambda_2
\end{equation}
for some $\lambda_1$, $\lambda_2$ all over the plane.
We consider the functional
\begin{equation}\label{functional}
I[U] = \int(\frac{1}{2}\nabla U\cdot\nabla U -\beta U + r \beta e^U)d^2z~.
\end{equation}
We start \cite{mcowen,troyanov} in the real pre-Hilbert space 
$H$ of $C^1$ functions $A$ with norm
\begin{equation}\label{sobolevnorm}
\langle A,A\rangle = \int \nabla A\cdot\nabla A d^2z+
\int A^2 \beta d^2z
\equiv (\nabla A,\nabla A)_1+(A,A)_\beta~.
\end{equation}
By $(A_1,A_2)_\beta$ we denote
the scalar product with the measure $\beta d^2z$ and with ${\cal L}^2_\beta$
the relative Hilbert space.
An simple computation gives 
\begin{equation}
\int (-A+r e^A )\beta d^2z
 \geq (1+\log \lambda_1) \int\beta d^2z
\end{equation}
showing that the functional $I$ is lower bounded in $H$.
We have also
\begin{equation}
I[0] = \int r \beta d^2z < \lambda_2\int \beta d^2z~.
\end{equation}

Defined
\begin{equation}
F=(r e^A-A)\beta~~~~{\rm and}~~~~
L = {\rm max}~(|\log\lambda_1|,|\log\lambda_2|)
\end{equation}
one has 
\begin{equation}\label{Fproperties}
\frac{\partial F}{\partial A} >0 ~~~~{\rm for}~~ A>L,~~~~
\frac{\partial F}{\partial A} <0 ~~~~{\rm for}~~ A<-L~.
\end{equation}

Thus given an $A$ one can construct an other $\tilde A\in C^1$ with
$|\tilde A|\leq 2 L$ 
and with the property
$I[\tilde A]\leq I[A]$. In fact consider a smooth always increasing
function $\sigma(x)$ with $\sigma(-\infty)=-2L$, 
$\sigma(\infty)=2L$, $\sigma(x)=x$ for $-L\leq x\leq L$ and elsewhere
$0<\sigma'<1$. Then with $\tilde A(z) = \sigma(A(z))$ we have
that $\int (-\beta A+r\beta e^{A})d^2z$ is not increased and also
\begin{equation}\label{tildabound}
\int \nabla \tilde A\cdot\nabla \tilde A~ d^2z=
\int (\sigma'(A))^2 \nabla A\cdot\nabla A ~d^2z\leq
\int \nabla A\cdot\nabla A d^2z~.
\end{equation}
This means that $\inf(I[A])$ can be computed on the subset $|A|\leq 2L$.
Note also that $\langle\tilde A,\tilde A\rangle\leq
\langle A, A\rangle$.
For $A\in H$, $|A|\leq 2L$ we have
\begin{equation}\label{Ibound}
I[A] =\frac{1}{2}\langle A,A\rangle - \frac{1}{2}(A,A)_\beta -\int
A\beta d^2z +\int r e^A \beta d^2z\geq\frac{1}{2}\langle A,A\rangle - 
2(L^2+L)(1,1)_\beta
\end{equation}
which shows that 
$\inf(I[A])$ is obtained using elements in the subset
of $H$ with $|A|\leq 2L$ and 
\begin{equation}\label{Mbound}
\frac{1}{2}\langle A,A\rangle \leq M+1
\end{equation}
being
\begin{equation}
M=I[0]+2(L^2+L)(1,1)_\beta~.
\end{equation}
In fact if $\frac{1}{2}\langle A,A\rangle >M+1$
we have
\begin{equation}
I[A]>M+1-2(L^2+L)(1,1)_\beta = I[0]+1
\end{equation}
and such $A$ has to be discarded in the search of ${\rm inf} ~I[A]$.

The subset ${\cal S}$ of $H$, and thus also of ${\cal L}^2_\beta$, 
given by  $\frac{1}{2}\langle A,A\rangle
\leq M+2$ and $|A|\leq 4L$ is relatively compact in
${\cal L}^2_\beta$, i.e. its closure is compact.

This is obtained by showing that the functions $\sqrt{\beta}A$ with
$A\in{\cal S}$
satisfy the three relative-compactness criteria \cite{HL}
in the usual ${\cal L}^2$ norm which we shall write as $||~~||$. To start
we notice that denoting by $\tau_h$ the operator which translates a
function by $h$ we have
\begin{equation}
\tau_hA(z)-A(z)=|h|\int_0^1 \hat h\cdot\nabla A(z+h \sigma)d\sigma
\end{equation}
and
\begin{equation}
\int|\tau_hA(z)-A(z)|^2 d^2z\leq 
|h|^2 \int_0^1d\sigma\int |\nabla A(z+\sigma h)|^2 d^2z
\leq |h|^2 ||\nabla A||^2
\end{equation}
i.e.
\begin{equation}\label{shift}
||\tau_hA-A||\leq |h|~||\nabla A||\leq |h|~\sqrt{\langle A,A\rangle}~.
\end{equation}

The criteria of relative compactness to be satisfied are \cite{HL}

1) The boundedness on ${\cal S}$ of $||\sqrt{\beta A}||$  which
is immediate due to $|A|\leq 4L$ and the 
integrability of $\beta$.

2) The uniformity on ${\cal S}$ of the limit 
$\lim_{R\rightarrow\infty} \int_{|z|>R}(\sqrt{\beta}A)^2 d^2z
=0$ which is true for  the same reason.

3) Finally we need to show that the limit 
\begin{equation}
\lim_{h\rightarrow 0} ||\tau_h
(\sqrt{\beta}A) - \sqrt{\beta}A||=0
\end{equation}
is uniform on ${\cal S}$. This is easily obtained from 
\begin{eqnarray}
&&||\tau_h(\sqrt{\beta}A) - \sqrt{\beta}A||\leq ||(\tau_h\sqrt{\beta})
(\tau_hA - A)||+||A(\tau_h\sqrt{\beta}- \sqrt{\beta})||\\
&&\leq||(\tau_h\sqrt{\beta})
(\tau_hA - A)||  +4L||\tau_h\sqrt{\beta}- \sqrt{\beta}||
\end{eqnarray}
and using eq.(\ref{shift}), $|A|\leq 4L$ and the integrability of $\beta$.

\bigskip

Construct now in ${\cal S}$ a 
sequence $A_m$ such that $\lim_{m\rightarrow\infty} I[A_m] =
\inf$. Actually due to the bounds given after eq.(\ref{tildabound}) 
and the bound of eq.(\ref{Mbound}) we can build this sequence with 
$|A_m|\leq 2L$ and $\frac{1}{2}\langle A_m,A_m\rangle \leq M+1$.

Due to the
relative-compactness of ${\cal S}$ in ${\cal L}^2_\beta$ we can extract a
sub-sequence $A_n$ such that it converges in ${\cal L}^2_\beta$ 
to some $U^*\in {\cal L}^2_\beta$. 
We shall have $|U^*|\leq 2L$ almost everywhere (a.e.).  

Then due to the continuity in ${\cal L}^2_\beta\cap\{|U|\leq 4L\}$ 
of $\int r e^{U}\beta d^2z$
we have 
\begin{equation}
\lim_{n\rightarrow\infty}I[A_n]= 
\lim_{n\rightarrow\infty}\int\frac{1}{2}\nabla A_n\cdot\nabla A_n
d^2z-\int U^*\beta d^2z+\int r e^{U^*}\beta d^2z ={\rm inf}~.
\end{equation}
Given a $\rho\in C_0^\infty$ the functions $A_n+\varepsilon\rho$ for
sufficiently small $|\varepsilon|$ belong to ${\cal S}$ and thus
\begin{eqnarray}
& &\lim_{n\rightarrow\infty}I[A_n+\varepsilon\rho] \\
&=&{\rm inf}+
\int\big(\frac{\varepsilon^2}{2}\nabla\rho\cdot\nabla\rho -
\varepsilon\Delta\rho U^* -
\varepsilon\rho\beta +r(e^{\varepsilon\rho}-1)
e^{U^*}\beta \big) d^2z \geq {\rm inf}~.
\end{eqnarray}
Using $e^x -1 -x \leq\frac{1}{2}x^2 e^{|x|}$ 
and the boundedness of $\rho$ we have  
\begin{equation}\label{linearrho}
0=\int(-\Delta\rho U^*-\rho\beta+r \rho e^{U^*}\beta)d^2z\equiv
(-\Delta\rho, ~U^*)_1+(\rho,re^{U^*}-1)_\beta
\end{equation}
for any $\rho\in C_0^\infty$.
Define now 
\begin{equation}\label{U1}
U_1(z) =
\frac{1}{4\pi}\int \log|z-z'|^2 \big(r(z')e^{U^*(z')}-1\big)\beta(z') d^2z'~.
\end{equation}
We have from (\ref{linearrho})
\begin{equation}
0=\int \Delta\rho (U^*-U_1)d^2z
\end{equation}
whose most general solution is, due to Weyl lemma \cite{FK}
\begin{equation}
U^* = U_1+ h~~~~~~~~a.e.
\end{equation}
with $h$ harmonic function. 
Thus we can now replace in (\ref{U1}) $U^*$ 
with $U_1+h$
obtaining
\begin{equation}\label{integreq}
U_1(z)= \frac{1}{4\pi}\int \log|z-z'|^2 
(r(z')e^{U_1(z')+h(z')}-1)\beta(z') d^2z~.
\end{equation}

Being $r$ and $U^*$ bounded and satisfying $\beta$ the bounds
(\ref{betabound1},\ref{betabound2},\ref{betabound3}),
eqs.(\ref{U1},\ref{integreq}) imply that $U_1$ is continuous with its first
and second derivatives and thus  $W\equiv U_1+h$
satisfies  
\begin{equation}
\Delta W=(r e^W-1)\beta
\end{equation}
which is eq.(\ref{newliouville}) and this concludes the existence proof 
for the sphere with elliptic singularities.

We can also determine the harmonic function $h$. Being
\begin{equation}
\int (re^{U^*}-1)\beta d^2z
\end{equation}
convergent, $U_1$ grows at infinity not faster than $\log z\bar z$ and the
boundedness of $U^*$ implies
\begin{equation}
h=c_1,~~~~~~~~\int (re^{U^*}-1)\beta d^2z = 0
\end{equation}
which fixes also the value of the constant $c_1$
\begin{equation}
e^{c_1}\int r\beta e^{U_1}d^2z = \int\beta d^2z~.
\end{equation}

\bigskip

In presence of parabolic singularities the positive 
function $\beta$ in addition
to the requirements
(\ref{betabound1},\ref{betabound2},\ref{betabound3})
is chosen in finite
domains $D_P$ around the parabolic singularity to be equal to
\begin{equation}
\beta = \frac{8}{|\zeta|^2 \log^2|\zeta|^2}
\end{equation}
with $\zeta = z - z_P$.

It is still possible to define a function $v$ such that
\begin{equation}
\Delta v= \beta,~~~~v\approx -2\log |z|^2 ~~{\rm for}~~z\rightarrow \infty~.
\end{equation}
First write
\begin{equation}
\phi_1=\sum_K (-2\eta_K)\log|z-z_K|^2-
\sum_P\log|z-z_P|^2.
\end{equation}
Then introduce \cite{lichtenstein} a smooth function $w_0$ with 
compact support which in each neighborhood $D_P$ of $z_P$ equals
\begin{equation}
-\log\log^2 |\zeta|^2
\end{equation}
and set
\begin{equation}\label{phi1parabolic}
v = \phi_1+w_0
+\frac{1}{4\pi}\int \log|z-z'|^2
\big(\beta(z') -\Delta w_0(z')\big)d^2z'~.
\end{equation}
We normalize then $\beta$ as to have
\begin{equation}
\frac{1}{4\pi}\int (\beta -\Delta w_0) d^2z =\frac{1}{4\pi}\int \beta d^2z
=-2+\sum_K2\eta_K+\sum_P 1
\end{equation}
which again is possible due to the topological inequality $\sum_K
2\eta_K+ \sum_P 1 >2$.

Then everything follows as in the case where only elliptic
singularities are present.

\section{\bf Extension to $g\geq 1$} 

We give now the extension of the previous results to the case of a
compact Riemann surface of genus $g\geq 1$. We deal first with 
$g \geq 2$. In this case  we can
represent such a surface by a standard fundamental polygon in the upper
$z$ plane \cite{FK}. 
This is a curvilinear $2g$-gon given by a sequence of arcs
$A_1 B_1 A_1^{-1} B_1^{-1}\dots A_n^{-1} B_n^{-1}$ where such arcs are pairwise
identified. The upper half plane is endowed with the metric
\begin{equation}
e^{\phi_B} dz\wedge d\bar z \frac{i}{2} =
\frac{8}{(z-\bar z)(\bar  z-z)}dz\wedge d\bar z \frac{i}{2}
\end{equation}
and we have
\begin{equation}
\Delta\phi_B=e^{\phi_B}~.
\end{equation}
Applying the Gauss-Bonnet relation \cite{FK}
\begin{equation}
\int K e^{\phi_B}dz\wedge d\bar z \frac{i}{2}=2\pi(2-2g)= 2\pi \chi_E
\end{equation}
and taking into account that the curvature $K$ is given by
\begin{equation}
K=-\frac{1}{2}\frac{\Delta\phi_B}{e^{\phi_B}}=-\frac{1}{2}
\end{equation}
we have for the area
\begin{equation}\label{area}
{\cal A} = \int e^{\phi_B}dz\wedge d\bar z \frac{i}{2}= 4\pi(2g-2)~.
\end{equation}
We split the field $\phi$ as
\begin{equation}\label{psi}
\phi = \phi_B+ \psi
\end{equation}
where $\psi$ obeys periodic boundary conditions. We have
\begin{equation}
\Delta\psi + e^{\phi_B} = e^{\psi+\phi_B}
\end{equation}
which, integrated, implies due to (\ref{area}) the topological restriction, 
\begin{equation}\label{topinequality}
\sum_K 2\eta_K+\sum_P 1 +2g-2>0~.
\end{equation}
In analogy to what done
for the sphere we set
\begin{equation}
\psi=U + v
\end{equation}
with, when in presence of only elliptic singularities,
\begin{equation}
v = 4\pi\sum_k -2\eta_k G(z,z_K) + \int G(z,z') \beta(z')d^2 z'
\end{equation}
where the positive function $\beta$ is chosen to satisfy
 around the singularities the properties given in the previous section
and with $\beta ~dz\wedge d\bar z$ periodic at the boundary.
$G(z,z')$ is the Green function of the Laplace-Beltrami operator
$e^{-\phi_B}\Delta$ on the fundamental polygon satisfying
\begin{equation}
4\partial_z\partial_{\bar z}G(z,z')=
\Delta G(z,z') = \delta(z,z') - \frac{e^{\phi_B(z)}}{{\cal A}}
\end{equation}
and ${\cal A}$ is given by eq.(\ref{area}).
Eq.(\ref{simpleliouville}) becomes 
\begin{equation}\label{reducedLiouvilleg}
\Delta U + \beta +\frac{e^{\phi_B}}{{\cal A}}\bigg(4\pi \sum_K 2\eta_K - \int\beta
d^2z\bigg)+e^{\phi_B}=e^{\phi_B} e^{v}e^\phi
\end{equation}
and we normalize the positive function $\beta$ as
\begin{equation}
\int\beta d z\wedge d\bar z \frac{i}{2} = 4\pi\big(\sum_K2\eta_K+2g-2\big)
\end{equation}
which is possible due to the topological inequality (\ref{topinequality}).
For the torus we simply employ $\phi_B=0$. Thus equation
(\ref{reducedLiouvilleg}) takes the form
\begin{equation}
\Delta U+\beta = e^{\phi_B}e^v e^U\equiv e^{\phi_B}r\beta e^U
\end{equation}
and the functional (\ref{functional}) becomes
\begin{equation}
I[U] = \int(\frac{1}{2}\nabla U\cdot\nabla U -\beta U + e^{\phi_B}r\beta 
 e^U)d^2z~.
\end{equation}
We proceed now as for the sphere. One starts from the real pre-Hilbert 
space $H$ of the $C^1$ functions satisfying periodic boundary 
conditions and with
norm (\ref{sobolevnorm}) where the integral is now extended to the
fundamental polygon. The relative-compactness of the subset ${\cal S}$ 
given by $|A|\leq 4L$, $\frac{1}{2}\langle A,A\rangle \leq M+2$ 
of $H$ is proven by
multiplying the periodic field in a neighborhood of the fundamental
polygon by a smooth positive function $\rho(z)$ which is $1$ inside
the polygon and vanishes outside such neighborhood and proceeding like
in the case of the sphere.

When also parabolic singularities are present one acts as in the case
of the sphere.

\bigskip
\section{\bf Uniqueness} 

We know that the solution of eq.(\ref{simpleliouville}) is locally
equivalent to the solution of the ordinary differential equation
in the complex plane (see e.g.\cite{CMS}) 
\begin{equation}\label{auxiliary}
f''(z)+ Q(z)f(z)=0
\end{equation}
which is known as auxiliary differential equation.
 
In a neighborhood of an elliptic singularity we have,
with $\zeta = u-u_K$ 
\begin{equation}\label{expressionK}  
\phi = -2\eta_K \log(\zeta\bar \zeta)-2\log [f(\zeta)\bar f(\bar
  \zeta)-\kappa^4(\zeta\bar\zeta)^{1-2\eta_K} g(\zeta)\bar g(\bar
  \zeta)]
\end{equation}  
where $f(\zeta)$ and $g(\zeta)$ 
are given by a locally convergent power expansions with non zero
constant terms and at infinity for the sphere 
$\phi= -2\log\zeta\bar\zeta + h(\frac{1}{\zeta},  
\frac{1}{\bar\zeta})$ with $h$ analytic function in the two variables.
Around parabolic singularities we have the expression \cite{CMS} 
\begin{equation}\label{expressionP}
\phi = -\log\zeta\bar\zeta - \log\log^2(\zeta\bar\zeta)-
2\log\bigg[g(\zeta)\bar g(\bar\zeta)+\frac{f(\zeta)\bar g(\bar\zeta)
+\bar f(\bar \zeta) g(\zeta)}{\log\frac{\zeta\bar\zeta}{\kappa^4}}\bigg]~.
\end{equation}  

Consider two solutions $\phi_1$ and $\phi_2$ of
eq.(\ref{simpleliouville}) satisfying 
eqs.(\ref{ellipticcondition},\ref{paraboliccondition},\ref{spherecond}). 
Then we have for the sphere, 
with $\partial f=\partial_zf dz$, 
$\bar\partial f=\partial_{\bar z}f d\bar z$
\begin{eqnarray}
0&\leq& \frac{i}{2}\int \partial(\phi_2-\phi_1)  
\wedge\bar\partial(\phi_2-\phi_1)= 
\frac{i}{2}\oint (\phi_2-\phi_1) \bar\partial(\phi_2-\phi_1)-
\frac{i}{2}\int (\phi_2-\phi_1)\partial\bar
\partial(\phi_2-\phi_1)\nonumber\\
&=&
0-\frac{1}{4}\int (\phi_2-\phi_1)
(e^{\phi_2}-e^{\phi_1})d^2z
\end{eqnarray}  
where the contour integral is around the singularities $u_K$ and $u_P$
and at
infinity and due to the behavior of $\phi_2-\phi_1$ given by
eq.(\ref{expressionK},\ref{expressionP}) it vanishes.
Thus we have
$\phi_2=\phi_1$.  For $g>0$ one acts exactly in the same way using
instead of $\phi$ the $\psi$ of eq.(\ref{psi}).
The usual uniqueness arguments \cite{picard1,picard2,poincare,lichtenstein} 
are more complicated
because they do not use the information about the non leading terms
appearing in the expansion of $\phi$ around the singularities 
and which we obtained from the expressions 
(\ref{expressionK},\ref{expressionP}).


\end{document}